# What indicators matter? The Analysis of Perception toward Research Assessment Indicators and Leiden Manifesto- The Case Study of Taiwan

Carey Ming-Li Chen*, Wen-Yau Cathy Lin**

* carey.mlchen@gmail.com
Science & Technology Policy Research and Information Center, National Applied Research Laboratories, 14F, No.106, Sec.2, Heping E. Rd., Taipei city 10636 (Taiwan)
Graduate Institute of Business Administration, National Taiwan University, No.1, Sec. 4, Roosevelt Rd., Taipei City 106 (Taiwan)

** wylin@mail.tku.edu.tw
Department of Information and Library Science, Tamkang University, No.151, Yingzhuan Rd., Tamsui Dist., New Taipei City 25137 (Taiwan)

**Introduction**

In 2015, five scientometricians published an article on the journal *Nature* to discuss some worrying phenomenon about the often ill applied in research evaluation of individual, institution, nation, or even the regional level (Hicks, Wouters, Waltman, de Rijcke, & Rafols, 2015). In fact, it is not the first time that academia discuss the misuse or abuse of bibliometric indicator, *San Francisco Declaration on Research Assessment (DORA)* shows up in 2012 and brings the debate of appropriateness of journal impact factor. Moreover, Higher Education Funding Council of England (HEFCE) published *The Metric Tide* in 2015 and starts to talk about what responsible metrics is. Nevertheless, is it the big improvement for academia in research evaluation?

In Taiwan, not only universities and research institutions, but also the major funding agency, Ministry of Science and Technology (MOST) adopts various quantitative indicators to evaluate researchers and faculty mainly, but their mindset about research evaluation has changed eventually and started to consider qualitative dimensions to assess the research performance instead of just counting "number." However, the discussions, arguments, and alarm about increasing misapplication of indicators have never been discontinued in this island. Despite that not every quantitative indicators are bibliometric indicators, the voice of against bibliometric indicators like JCR-Journal Impact Factor (JIF) or citation-based indicators has been getting louder recently. Some scholars even declare that over-emphasizing on SCI and SSCI is the source of evil and cause the whole academia only cares about JIF instead of research quality. However, is it fair to be blamed on the bibliometric indicators entirely? By examining the incentive policy in universities in Taiwan, it is found that misuse of terminology of journal impact factor happens a lot, and not to mention the lack of understanding of bibliometric indicators. From the perspective of a steady evaluation mechanism, all of the relevant people, including evaluated, assessors, and research administrators ought to know the indicators well. Should this be the case, the dispute will not be so much. Under this circumstance, our study aims to fulfil the curiosity about the perception toward the concept of bibliometrics and indicators in Taiwan academia. After all, it is important to investigate the awareness at first before initiating the design of guideline of research evaluation locally (Chen & Lin, 2017).  At the meantime, to see whether it is appropriate to adjust the content of Leiden Manifesto to fit the local context, this study aims to understand the Taiwanese researchers' perception toward ten principles.  This action is considered to be the approach to introduce Leiden Manifesto to Taiwan academia as well.

**Research Methods**
**Questionnaire Design**
The questionnaire consists of four parts.  The first section is to examine the perception toward the concept of bibliometrics, bibliometric indicators and important works about the reflection on use of research assessment indicators.  The respondents are asked to evaluate their understandings to the bibliometric indicators.  To compare the survey result with previous work (Rousseau & Rousseau, 2017), this study lists several common bibliometric indicators, such as JCR's journal impact factor, SCImago Journal Ranking, SNIP, Eigenfactor, 5-year synchronous journal impact factor, relative citation ratio (RCR) and *h*-index.  The respondents are required to answer whether they heard the indicators and are familiar with the definition or formula of calculation.  The first section also lists questions to examine whether the respondents are aware of the three important publications in bibliometrics and research evaluation including *DORA*, *Leiden Manifesto* and *The Metric Tide* (OECD, 2016).

The second section is to understand the scenarios that the researchers use bibliometric indicators.  The scenarios of using bibliometrics include grant proposal review, promotion evaluation, recruitment, institution or universities assessment, searching for journal articles and journal manuscript review.  This study also aims to know whether being at the different circumstances affects researchers' perception toward to indicators, hence the questionnaire lists the question to ask the respondents to examine their "relationship" with bibliometric indicators to see whether they belong to the developers of bibliometric indicators, assessors, or assessees.

The third section is to know the perception toward *Leiden Manifesto* among Taiwanese researchers.  The questionnaire lists ten principles from *Leiden Manifesto* and utilizes Likert Scale to assess the degree of agreement.  From 1 to 5 is strongly disagree to strongly agree.  The following question is to ask the respondents to choose three principles which are the most important principles to them.  The final question is not the required but optional to ask the respondents to provide their opinions about the utilization of bibliometric indicators in grant proposal review, recruitment, and research evaluation.  The link of full-text of *Leiden Manifesto* and the translations of Traditional Chinese version are provided and listed on the beginning of this section in the questionnaire.  In fact, this action aims to promote the content of *Leiden Manifesto* to Taiwanese researchers as well.  The respondents can have more comprehensive understanding about the bibliometric indicators and research evaluation via filling this questionnaire.

The final section of this questionnaire is demographics.  The items include academic position, type of affiliation, age distribution, gender, the highest education level and their major discipline based on the classification system of Ministry of Science and Technology in Taiwan.

**Data Collection**
This study applied online survey and the respondents were invited to answer the questionnaire. The invited respondents were selected from the authors of publications indexed in Science Citation Index (SCI) and Social Science Citation Index (SSCI) in Web of Science. After excluding e-mails with foreign country domain name, a total of 8,514 e-mail invitations were sent. The survey period lasts two weeks from the final week of February to March.

**Research Results**
Overall, a total of 421 respondents answered the survey. After excluding the invalid answers, the number of valid respondents was 417. About 80% of respondents was male, and most of them were 40~44 years old (22.54%). 41.97% of respondents served in public universities and most of their position was professor (45.08%). The distribution of disciplines was 35.73% from life science, 29.50% from engineering and technology, 17.50% from humanities and social science, 14.39% from natural science, and 2.88% from science education.

*Awareness toward bibliometric indicators*
The first section of questionnaire was designed to understand the respondents' awareness about bibliometric indicators. First of all, 37.41% of respondents reported themselves not hearing *informetrics* before, only 3.12% of them declared them fully understand the content of *informetrics*. Moreover, over 65% of respondents did not hear bibliometric indicators before, only 3.36% of them considered themselves understand completely the content of bibliometric indicators (Figure 1).

Figure 1: The level of awareness toward informetrics and bibliometric indicators.

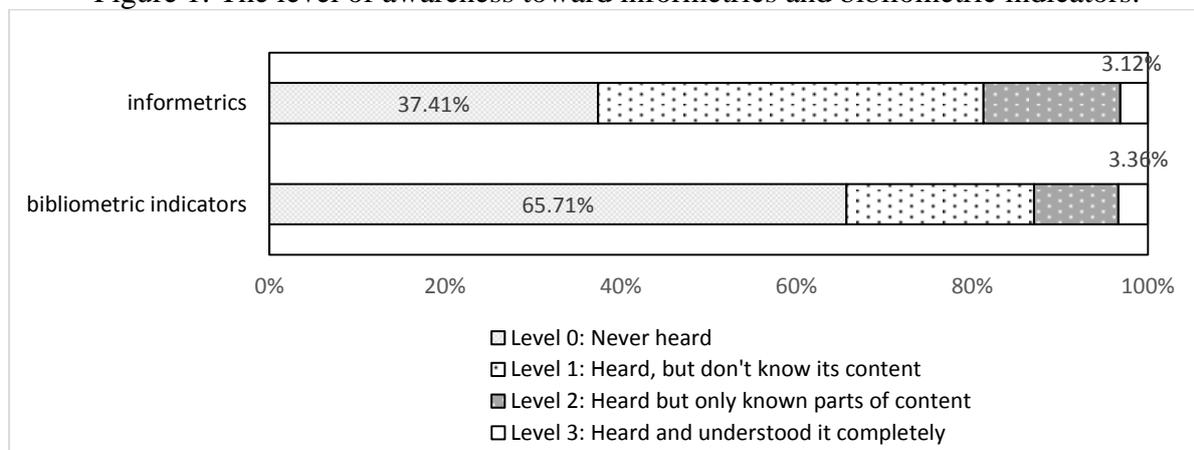

In the following questions, respondents are asked to evaluate whether they know several bibliometric indicators, the awareness level is defined to four level. The result is shown in Figure 2. Surprisingly, although over 65% of respondents thought themselves not hearing bibliometric indicators before, only 0.48% of respondents reported them not hearing JIF. In other words, the almost all of the researchers in Taiwan have ever heard of JIF, and 62.35% of them even considered themselves completely understand what it is JIF. The other indicator related to journal ranking is SCImago Journal Rank, 24.70% of respondents thought themselves understand it as well. Nevertheless, other bibliometric indicators seem not to be recognized very well by the researchers in Taiwan, especially for SNIP, Eigenfactor and relative citation ratio (RCR), only 7.67% of respondents committed themselves fully understand RCR. On the other hand, the indicator applied to individual level evaluation mostly is recognized by Taiwanese researchers better, 28.78% of respondents heard and understood what *h*-index it is completely.

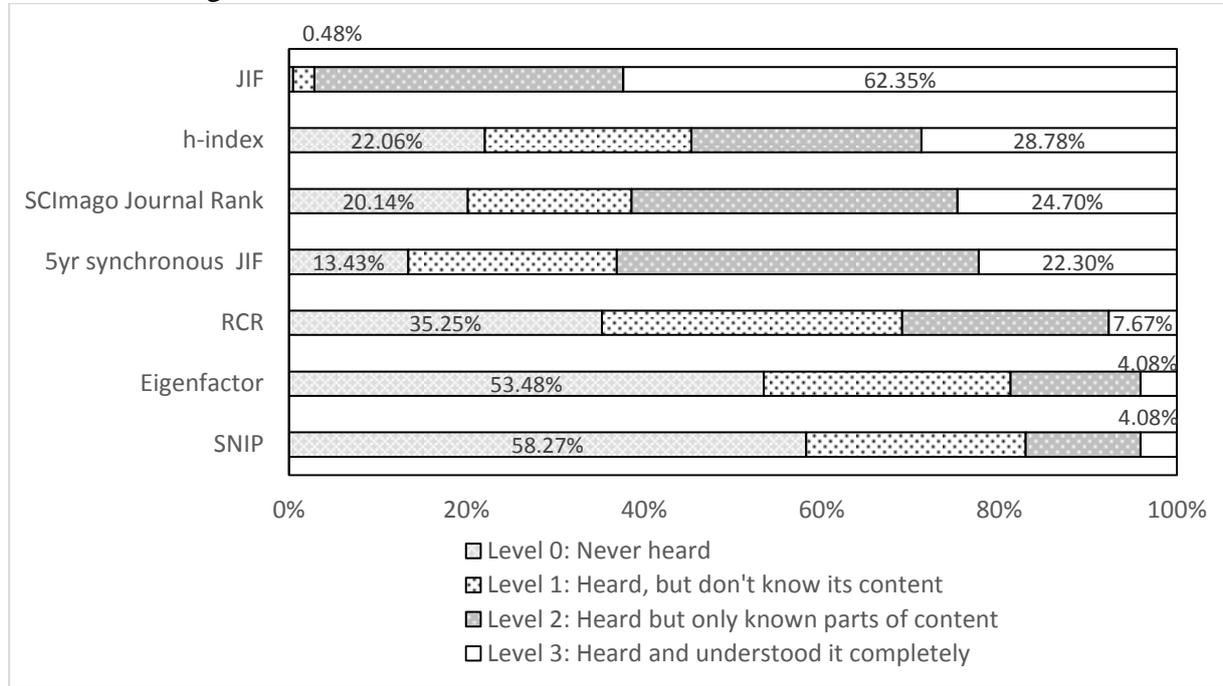

Figure 2: The level of awareness toward the bibliometric indicators.

*Awareness toward important works regarding to the reflection of bibliometric indicators*
This study also aims to examine the respondents' awareness toward three important works regarding to the reflection of bibliometric indicators. The result is shown in Figure 3. The majority of Taiwanese researchers have not heard of *DORA*, *Leiden Manifesto* and *The Metric Tide*. It shows that these works did not be paid attention by Taiwan academia and it needs effort to let researchers aware these documents and then it is possible to arisen the right concepts of use of bibliometric indicator in research evaluation. The work which is recognized by Taiwanese researchers the most is *DORA*, 28.54% of the respondents have heard this document, but only 2.64% of them fully understand its content.

Figure3. The level of awareness toward important works regarding to the reflection of bibliometric indicators.

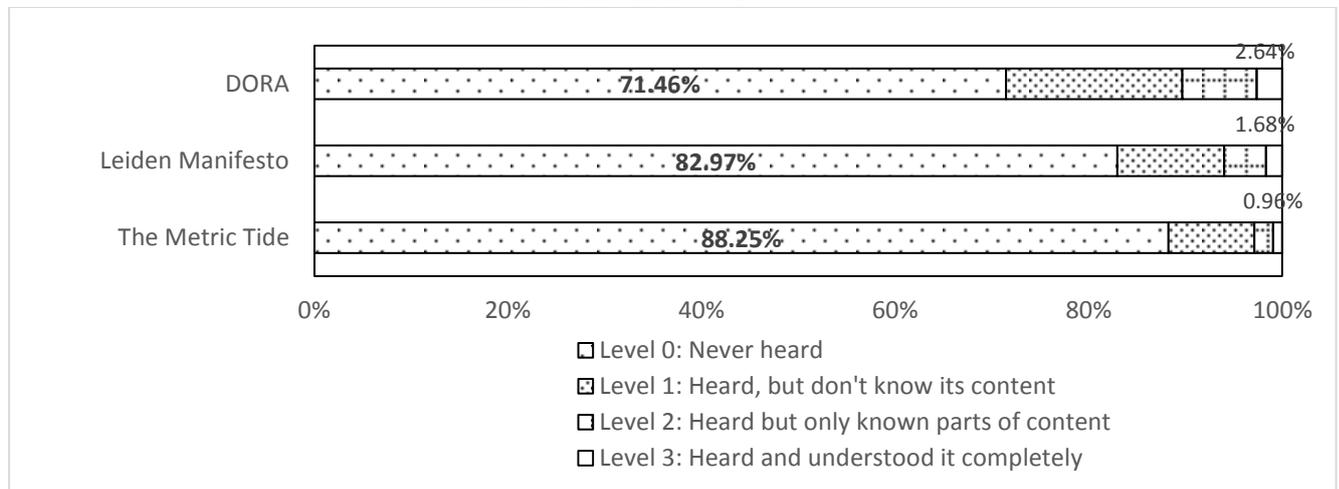

*The timing of using bibliometric indicators*
The survey conducts what scenarios are researchers use bibliometric indicators. The result is presented in Figure 4. The researchers use bibliometric indicators when they review grant proposal at most. Besides, when evaluating the researcher promotion, the bibliometric indicators can be used as well. Another main scenario when researchers applying bibliometric indicators are searching for journal articles they needed. It means that researchers may use bibliometric indicator as basis to assist their judgement of article quality. This study asks the respondents to report their relationship with bibliometric indicators, and 73% of respondents considered them as the ones who are evaluated by bibliometric indicators. Less than half of respondents (45%) claimed themselves would use bibliometric indicator to evaluate others. Only 64 respondents said they were doing the work related to indicator development.

Figure4. The scenarios of use bibliometric indicators.

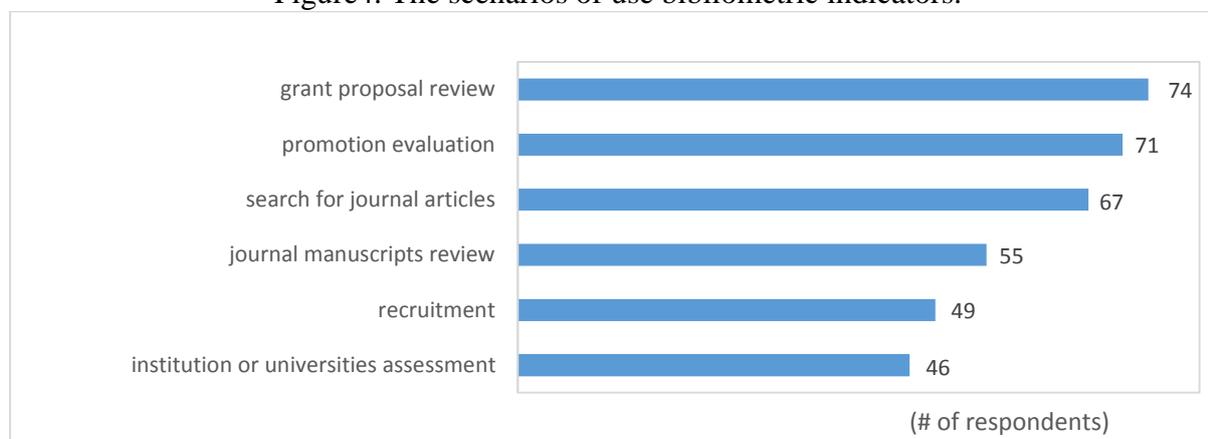

*Perception toward the ten principles of Leiden Manifesto*
To letting more people aware *Leiden Manifesto*, this study aims to ask the respondents to review ten principles and check the degree of agreement. We hope that this survey could be considered as the approach to disseminate the correct concept of indicators and research evaluation as well. Generally the respondents agree the statement of principles in *Leiden Manifesto*, almost each principles got four points in Likert scale. The highest one is Principle 6, "*Account for variation by field in publication and citation practices*" by getting 4.42. The second highest one is Principle 4, "*Keep data collection and analytical processes open, transparent and simple*" (4.39). The third one is Principle 8 "*Avoid misplaced concreteness and false precision*" (4.28). More details are shown in Table 1. The respondents are also asked to check three the most important principles of *Leiden Manifesto* in their perspectives, and the result is shown in Figure 5. Principle 6 is considered to be the most important principle in research evaluation since over half of the respondents picked. Principle 4 earns the second place, and this result is consistent with Table 1. However, principle 1 "*Quantitative evaluation should support qualitative, expert assessment*" is placed the third one, but it only got 3.95 in 5-point Likert scale measurement.

Based on these results, it is found that the respondents care whether research evaluation considers the different characteristics of fields, the transparent of data collection, and the precision of bibliometric indicators. It indicated that Taiwanese researchers may worry about research evaluation jumps to wrong conclusion by misusing indicators instead of considering the normalization. However, comparing with the previous result of awareness of bibliometric indicators, few researchers realize the definition of relative citation ratio which applied field-normalization concept, it perhaps implies that Taiwanese researchers aware the importance of field normalization under the certain context in research evaluation, but they do not realize that the bibliometric indicators already evolve, and that's why they are worried about abuse of bibliometric indicators, and even complain about the indicators themselves.

Table 1. The perception toward ten principles of Leiden Manifesto

| **Principles** | **Mean** | **S.D.** |
|---|---|---|
| 1. Quantitative evaluation should support qualitative, expert assessment. | 3.95 | 0.94 |
| 2. Measure performance against the research missions of the institution, group or researcher. | 4.00 | 0.93 |
| 3. Protect excellence in locally relevant research. | 4.04 | 0.95 |
| 4. Keep data collection and analytical processes open, transparent and simple. | 4.39 | 0.77 |
| 5. Allow those evaluated to verify data and analysis. | 4.25 | 0.80 |
| 6. Account for variation by field in publication and citation practices. | 4.42 | 0.79 |
| 7. Base assessment of individual researchers on a qualitative judgement of their portfolio. | 4.11 | 0.83 |
| 8. Avoid misplaced concreteness and false precision. | 4.28 | 0.78 |
| 9. Recognize the systemic effects of assessment and indicators. | 4.16 | 0.79 |
| 10. Scrutinize indicators regularly and update them. | 4.26 | 0.80 |

Note: The measurement applies 5-point Likert scale. From 1 to 5 is strongly disagree to strongly agree.

Figure 5. The most important principles of Leiden Manifesto

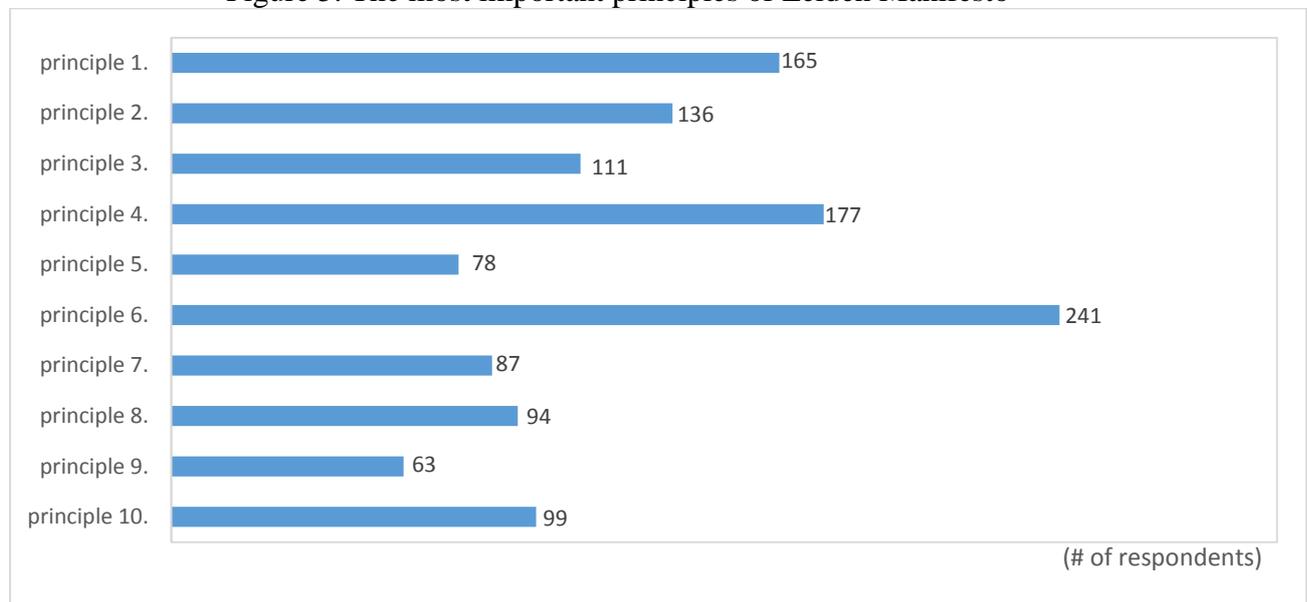

(# of respondents)

*Does research field make it difference?*
About the level of awareness toward the bibliometric indicators, the result is listed in Table 2. Except journal impact factor ($\chi^2=20.80$, p-value<0.1) and *h*-index ($\chi^2=20.08$, p-value<0.1), there is no significant difference in recognition of other five bibliometric indicators among the researchers from different fields. It means that the lack of understanding toward bibliometric indicator is universal phenomenon in Taiwan. However, the researchers from life science have highest percentage in level 3 in journal impact factor, 75.17% of them realize this indicator very well, and no respondents from life science declare them not hearing journal impact factor. This result implies that the use of journal impact factor did play an important role in research evaluation of life science field, and it illustrates why DORA is drafting by the experts and scholars in life science. About *h*-index, it is another well-known indicator, except the researchers from humanities and social science, only 12.32% of respondents heard and understood completely it, other four groups of researchers thought themselves understand it and are quite familiar with its definition.

Table 2. The percentage of different levels of awareness toward the bibliometric indicators among the groups

|  | Life Sci (n=149) | Eng (n=123) | HSS (n=73) | Natural Sci (n=60) | Sci Edu (n=12) | $\chi^2$ |
|---|---|---|---|---|---|---|
| Journal Impact Factor |  |  |  |  |  | 20.80* |
| Level 0 | 0.00 | 0.81 | 0.00 | 1.67 | 0.00 |  |
| Level 1 | 1.34 | 3.25 | 2.74 | 3.33 | 0.00 |  |
| Level 2 | 23.49 | 37.40 | 43.84 | 43.33 | 50.00 |  |
| Level 3 | 75.17 | 58.54 | 53.42 | 51.67 | 50.00 |  |
| 5yr Journal Impact Factor |  |  |  |  |  | 9.48 |
| Level 0 | 16.11 | 11.38 | 13.69 | 11.66 | 8.33 |  |
| Level 1 | 18.12 | 31.71 | 23.29 | 20.00 | 25.00 |  |
| Level 2 | 42.95 | 34.15 | 42.47 | 46.67 | 41.67 |  |
| Level 3 | 22.82 | 22.76 | 20.55 | 21.67 | 25.00 |  |
| SCImago Journal Rank |  |  |  |  |  | 17.35 |
| Level 0 | 20.80 | 14.63 | 20.55 | 30.00 | 16.66 |  |
| Level 1 | 15.44 | 17.89 | 28.76 | 18.33 | 0.00 |  |
| Level 2 | 39.60 | 40.65 | 27.40 | 31.67 | 41.67 |  |
| Level 3 | 24.16 | 26.83 | 23.29 | 20.00 | 41.67 |  |
| SNIP |  |  |  |  |  | 13.70 |
| Level 0 | 58.39 | 57.72 | 61.64 | 56.67 | 50.00 |  |
| Level 1 | 30.20 | 17.89 | 23.29 | 26.67 | 25.00 |  |
| Level 2 | 8.05 | 17.89 | 12.33 | 15.00 | 16.67 |  |
| Level 3 | 3.36 | 6.50 | 2.74 | 1.66 | 8.33 |  |
| Eigenfactor |  |  |  |  |  | 17.14 |
| Level 0 | 56.38 | 55.29 | 56.16 | 45.00 | 25.00 |  |
| Level 1 | 30.87 | 27.64 | 16.44 | 31.67 | 41.67 |  |
| Level 2 | 10.74 | 13.82 | 20.55 | 16.67 | 25.00 |  |
| Level 3 | 2.01 | 3.25 | 6.85 | 6.66 | 8.33 |  |
| RCR |  |  |  |  |  | 6.99 |
| Level 0 | 36.24 | 34.96 | 36.98 | 36.67 | 8.33 |  |
| Level 1 | 32.22 | 34.15 | 32.88 | 36.67 | 41.67 |  |
| Level 2 | 25.50 | 22.76 | 20.55 | 18.33 | 41.67 |  |
| Level 3 | 6.04 | 8.13 | 9.59 | 8.33 | 8.33 |  |
| *h*-index |  |  |  |  |  | 20.08* |
| Level 0 | 21.48 | 17.08 | 32.88 | 23.33 | 8.33 |  |
| Level 1 | 22.82 | 25.20 | 27.40 | 18.33 | 8.33 |  |
| Level 2 | 22.82 | 25.20 | 27.40 | 30.00 | 41.67 |  |
| Level 3 | 32.88 | 32.52 | 12.32 | 28.34 | 41.67 |  |

Note 1: * indicates p-value<0.1; * indicates p-value<0.05; *** indicates p-value<0.01
Note 2: Level 0: Never heard; Level 1: Heard, but don't know its content; Level 2: Heard but only known parts of content; Level 3: Heard and understood it completely.
Abbreviation: Life Sci: Life Science; Eng: Engineering and Technology; HSS:Humanities and Social Sciences; Natural Sci: Natural Science and Sustainable Development; Sci Edu:Science Education

Table 3 listed the result about researchers' awareness toward *DORA*, *Leiden Manifesto* and *The Metric Tide* among different fields. Only significant difference in DORA exists among five groups of researchers ($\chi^2$=27.60, p-value <0.01). The result indicates that Taiwanese researchers generally do not aware the existence of *Leiden Manifesto* and *The Metric Tide*, even if they have heard of these documents, they are not familiar with the contents. In comparison, *DORA* has higher visibility to Taiwanese researchers. Not surprisingly, *DORA* earns more attention from the researchers in life science. The researchers from humanities and social science have little knowledge about *DORA*, only about 17% of respondents declared themselves hearing of *DORA* before.

Table 3. The percentage of different levels of awareness toward the important works about the reflection of bibliometric indicators among the groups

|  | Life Sci (n=149) | Eng (n=123) | HSS (n=73) | Natural Sci (n=60) | Sci Edu (n=12) | $\chi^2$ |
|---|---|---|---|---|---|---|
| DORA |  |  |  |  |  | 27.60*** |
| Level 0 | 64.43 | 74.80 | 83.56 | 73.33 | 41.67 |  |
| Level 1 | 20.13 | 16.26 | 10.96 | 18.33 | 58.33 |  |
| Level 2 | 10.07 | 8.13 | 4.11 | 6.67 | 0.00 |  |
| Level 3 | 5.37 | 0.81 | 1.37 | 1.67 | 0.00 |  |
| Leiden Manifesto |  |  |  |  |  | 11.82 |
| Level 0 | 79.19 | 83.74 | 87.67 | 88.33 | 66.67 |  |
| Level 1 | 12.75 | 11.38 | 6.85 | 6.67 | 33.33 |  |
| Level 2 | 5.37 | 4.07 | 4.11 | 3.33 | 0.00 |  |
| Level 3 | 2.69 | 0.81 | 1.37 | 1.67 | 0.00 |  |
| The Metric Tide |  |  |  |  |  | 7.22 |
| Level 0 | 86.58 | 88.62 | 89.04 | 91.66 | 83.33 |  |
| Level 1 | 10.74 | 9.76 | 5.48 | 5.00 | 16.67 |  |
| Level 2 | 2.01 | 0.81 | 4.11 | 1.67 | 0.00 |  |
| Level 3 | 0.67 | 0.81 | 1.37 | 1.67 | 0.00 |  |

Note 1: * indicates p-value<0.1; * indicates p-value<0.05; *** indicates p-value<0.01
Note 2: Level 0: Never heard; Level 1: Heard, but don't know its content; Level 2: Heard but only known parts of content; Level 3: Heard and understood it completely.
Abbreviation: Life Sci: Life Science; Eng: Engineering and Technology; HSS:Humanities and Social Sciences; Natural Sci: Natural Science and Sustainable Development; Sci Edu:Science Education

*Perception toward the ten principles of Leiden Manifesto among groups of researchers in different disciplines*

This study also aims to understand whether the principles in *Leiden Manifesto* is able to be considered as general guideline and suits in every fields. Hence, this study conducts statistical analysis to see any significant difference exists. Since the 5-point Likert scale is ordinal and the responses do not follow the normal distribution, this study utilized Kruskal-Wallis test to examine. Based on the result listed in Table 4, Principle 2, 8, 9, 10 have significant difference among five groups of researchers. Other six principles have no significant difference, hence, they may be considered to be the universal principles in research evaluation. Especially the Principle 6 can almost be considered the consensus among different groups of researchers, the median number is 5, and it means that all researchers strongly agree that statement except for the researchers from science education (median=4.5). In Principle 2 "*Measure performance against the research missions of the institution, group or researcher*", although the median number is 4.00, the respondents from engineering and technology has higher share of disagree (8.13%) and strongly disagree (4.07%), and it means that the researchers may have disagreement about Principle 2 in that field. About the Principle 8 and the Principle 10, there exist certain level of disagreement. The researchers from humanities and social science and science education strongly agree that it is important to avoid misplace concreteness and false precision and often update the indicators when assessment compared to the researchers who are from the fields related to STEM.

Table 4. The median number of the perception toward ten principles of Leiden Manifesto among the groups

|  | Life Sci (n=149) | Eng (n=123) | HSS (n=73) | Natural Sci (n=60) | Sci Edu (n=12) | Kruskal-Wallis test $\chi^2$ |
|---|---|---|---|---|---|---|
| Principle 1. | 4.00 | 4.00 | 4.00 | 4.00 | 4.00 | 4.34 |
| Principle 2. | 4.00 | 4.00 | 4.00 | 4.00 | 4.00 | 10.12** |
| Principle 3. | 4.00 | 4.00 | 4.00 | 4.00 | 4.00 | 6.51 |
| Principle 4. | 5.00 | 5.00 | 5.00 | 4.00 | 5.00 | 1.80 |
| Principle 5. | 4.00 | 4.00 | 4.00 | 4.00 | 4.50 | 2.92 |
| Principle 6. | 5.00 | 5.00 | 5.00 | 5.00 | 4.50 | 2.58 |
| Principle 7. | 4.00 | 4.00 | 4.00 | 4.00 | 5.00 | 2.81 |
| Principle 8. | 4.00 | 4.00 | 5.00 | 4.00 | 5.00 | 8.87* |
| Principle 9. | 4.00 | 4.00 | 4.00 | 4.00 | 4.50 | 11.38** |
| Principle 10. | 4.00 | 4.00 | 5.00 | 4.00 | 5.00 | 18.18*** |

Note: * indicates p-value<0.1; * indicates p-value<0.05; *** indicates p-value<0.01
Abbreviation: Life Sci: Life Science; Eng: Engineering and Technology; HSS:Humanities and Social Sciences; Natural Sci: Natural Science and Sustainable Development; Sci Edu:Science Education

**Conclusions**
This results from this study demonstrate that evoking the right concept of use of bibliometric indicators and research evaluation has a long way to go. The lack of recognition of bibliometric indicators exists in Taiwan academia. Generally speaking, researchers may hear of the certain indicators, but they are not familiar with its definition and calculation process. Only JIF and *h*-index are considered as well-known indicators. Hence, when talking about the important works about evoking the importance of responsible metrics, *DORA* has higher visibility, few researchers in Taiwan academia aware the existence of *Leiden Manifesto*, not to mention *The Metric Tide*. Although *Leiden Manifesto* as publication in journal article is considered as the better way to disseminate the concept and earns more citations (Chen & Lin, 2017), the survey result indicates that Taiwanese researchers are not aware that. Hence, this study aims to promote the right concepts of research evaluation by conducting questionnaire survey and hope to analyse the perception toward ten principles in Taiwan academia.

The results suggest that the ten principles can be considered the universal guideline in research evaluation since most of Taiwanese researchers agree the contents of ten principles. Especially for the rinciple 6 "*Account for variation by field in publication and citation practices*" has less room of opinion diversity. However, it is interesting to compare the result of recognition of relative citation ratio, only few researchers have fully understood the definition. This result indicates that scientometricians should need to make more effort to disseminate the concept of field-normalization in bibliometric indicators. The researchers do have understanding about the importance of comparison on the same basis, at the meantime, they may use the inappropriate indicators just because of lacking of enough knowledge on the variety of indicators. Hence, it is important to initiate that education of informetrics to all of the stakeholders in research evaluation. Only having enough and accurate knowledge, the misuse and abuse of bibliometric indicators may possibly not happen again, and the bibliometric analysis is able to turn to contextualization-based analysis in the future.